\documentclass[preprint,12pt]{elsarticle}
\usepackage{amssymb}
\usepackage{amsthm}
\usepackage{lineno}
\usepackage{siunitx}
\usepackage{multirow}
\usepackage{natbib}
\usepackage{caption}
\usepackage{mathtools}
\usepackage{breqn}
\usepackage{amsmath}
\usepackage{color}
\usepackage{ulem}
\usepackage {cancel}
\usepackage{array}
\usepackage{amssymb}
\usepackage{caption}
\usepackage{url}
\usepackage{subcaption}
\usepackage{graphicx}

\journal{Journal of Sound and Vibration}
\biboptions{longnamesfirst,square,sort&compress}

\newcommand{\bhline}[1]{\noalign{\hrule height #1}}

\begin{document}
	\begin{frontmatter}
		\title{Sound Source Localization for a Source inside a Structure using Ac-CycleGAN}
		\author[1]{Shunsuke KITA\corref{cor1}}\ead{kitas@orist.jp}
		\author[1]{Choong Sik Park\corref{cor2}}\ead{park@orist.jp}
		\author[3]{Yoshinobu KAJIKAWA\corref{cor3}}\ead{kaji@kansai-u.ac.jp}
		\cortext[cor1]{Corresponding author}
		\address[1]{Osaka Research Institute of Industrial Science and Technology 2-7-1 Ayumino, Izumi-shi, Osaka, 594-1157 Japan}
		\address[3]{Faculty of Engineering Science, Kansai University 3-3-35 Yamate, Suita-shi, Osaka, 564-8680 Japan}
		\begin{abstract}
			We propose a method for sound source localization (SSL) for a source inside a structure using Ac-CycleGAN under unpaired data conditions.
			The proposed method utilizes a large amount of simulated data and a small amount of actual experimental data to locate a sound source inside a structure in a real environment.
			An Ac-CycleGAN generator contributes to the transformation of simulated data into real data, or vice versa, using unpaired data from both domains.
			The discriminator of an Ac-CycleGAN model is designed to differentiate between the transformed data generated by the generator and real data, while also predicting the location of the sound source.
			Vectors representing the frequency spectrum of the accelerometers (FSAs) measured at three points outside the structure are used as input data and the source areas inside the structure are used as labels.
			The input data vectors are concatenated vertically to form an image.
			Labels are defined by dividing the interior of the structure into eight areas with one-hot encoding for each area.
			Thus, the SSL problem is redefined as an image-classification problem to stochastically estimate the location of the sound source.
			We show that it is possible to estimate the sound source location using the Ac-CycleGAN discriminator for unpaired data across domains.
			Furthermore, we analyze the discriminative factors for distinguishing the data.
			The proposed model exhibited an accuracy exceeding 90\% when trained on 80\% of actual data (12.5\% of simulated data).
			Despite potential imperfections in the domain transformation process carried out by the Ac-CycleGAN generator, the discriminator can effectively distinguish between transferred and real data by selectively utilizing only those features that generate a relatively small transformation error.
		\end{abstract}
	\end{frontmatter}
\section{Introduction}
	Various sound source localization (SSL) methods have been proposed to date to estimate the location of structure-generated noise.
	The methods have been used in the development of automobiles, machineries, and home appliances \cite{knapp, carter}.
	For example, the noise from a car engine cooling system and the chattering noise from the vibrations of mechanical products can be visualized using single or array microphones and cameras \cite{AMOIRIDIS2022116534}.
	Thus, to reduce noise, it is essential to visualize the sound sources originating from the vibrations of structures, and it is also an important aspect of design in many engineering applications \cite{MARTINS2023117891}.
	SSL systems depend on the estimation of time difference of arrival (TDOA) in either or both time and frequency domains and assume microphone independence because of the direct sound.
	Deep learning approaches have been applied to SSL methods in recent years \cite{https://doi.org/10.48550/arxiv.2109.03465}.
	Nevertheless, existing methods are limited in their ability to estimate the position of acoustic radiation from the external surface of a structure, and cannot indirectly estimate the position of sound generated by a source within the structure from the outside
	Estimating interior noise sources from outside the structure is important because it is a means of providing a fundamental solution for lowering the noise levels of products.
	
	In recent years, there has been growing interest in research on indirectly estimating the position of sound sources inside a duct from outside the duct \cite{gao2023localization}.
	However, when the effects of vibrations of structures cannot be ignored, such as when a flat panel vibrates, the modal vibration response of the structure depends on the direction of arrival (DOA) of the wave, varying in response to the incident sound pressure wave \cite{DIPASSIO2023117671}.
	In such cases, TDOA-based methods cannot be applied because the sound propagating through the structure is indirect, and independence at the observation point cannot be guaranteed.
	To solve this problem, we propose novel method to locate a sound source inside the structure.
	The method uses a model trained using simulation data \cite{kita2021fundamental}.
	In this study, we show that the source position can be identified from the measured data by measuring the vibrations of the structure originating from the sound source within the structure, using acceleration sensors, from both a real environment and a simulation.
	
	In general, the performance of models pre-trained using simulation data degrades significantly when tested with real data because of the discrepancy between the distribution of simulation and real data \cite{poschadel2021direction, he2019adaptation, takeda2017unsupervised, takeda2018unsupervised, he2021neural}.
	This is called a ``domain shift" in the context of machine learning.
	This issue is faced not only in SSL but also in other fields, and it is due to differences in the population distributions from which the data are sampled\cite{shimodaira2000improving, moreno2012unifying, datashift}.
	Numerous studies have been conducted on ``transfer learning" or ``domain adaptation" to address the performance degradation caused by the differences in population distributions, and most of the studies have focused on image classification or semantic segmentation as the target task\cite{tan2018survey, wang2018deep}.
	The datasets used in these studies contain large amounts of image data, and the problem is set under a condition in which the labels of the target domain are unknown (unsupervised domain adaptation).
	The main approaches used to solve this issue include ``domain-invariant feature larning,” ``domain mapping,” and ``target discrimination" \cite{tzeng2017adversarial, hoffman2018cycada, saito2018maximum, ajakan2014domain, laradji2020m, wilson2020survey}.
	
	To solve the problem of SSL inside a structure, a large amount of simulation data can be generated; however, real data are difficult to obtain.
	In view of the above, we propose a domain-mapping-based method for locating a sound source inside a structure.
	The method is applicable not only to classification but also to regression problems for a small real dataset\cite{kita_semi, kita2023}.
	To address the performance degradation in SSL owing to domain shift, the proposed SSL method incorporates a domain transfer (DT) model to bridge the discrepancy between the simulation and real data.
	In this approach, the distribution of real data is made closer to that of simulated data via DT model transformation.
	The DT model transformation allows an SSL model that is trained using simulated data to adapt to real data.
	
	However, this approach has two objective functions to minimize: ``Categorical cross entropy" or ``Root mean squared error (RMSE)" set for the SSL model and the ``Mean squared error (MSE)" set for the DT model.
	Because of the different objective functions, the efficiency of the method in transforming the data does not directly contribute to the performance of the SSL model.
	In other words, there is no guarantee that the data transformed by the DT model will be embedded within the desired class of the SSL-model decision boundaries.
	Furthermore, this method requires paired data from both domains, which are expensive to collect.
	To solve this problem, we propose an SSL method that is applicable to unpaired datasets and utilizes a DT model combined with an SSL model.
	
	CycleGAN \cite{zhu2017unpaired} is a model that does not require paired data, in contrast to pix2pix\cite{isola2017image}.
	CycleGAN allows bidirectional domain transformation through adversarial training \cite{goodfellow2014generative} between two generators and their corresponding discriminators.
	It is widely used in the fields of image and natural language processing.
	Conditional CycleGAN (C-CycleGAN) uses the structure of conditional GAN (C-GAN) \cite{mirza2014conditional} to CycleGAN \cite{yu2019improving, tang2019expression}.
	C-CycleGAN aims to improve the performance of generating the target class data by inputting a conditional vector into the generators and discriminators.
	Although few studies have been conducted on acoustic signal processing, it is beginning to be used in the field of speech conversion.
	In these studies, mel-cepstrum coefficients, aperiodicity, and logarithmic fundamental frequencies were extracted from speech waveforms and formatted as image inputs for a C-CycleGAN model \cite{yook2018voice, lee2020many}.
	However, unlike a C-GAN model, auxiliary classifier GAN (AC-GAN) \cite{odena2017conditional} models add classification units to the discriminator's output layer.
	AC-GAN has a structure in which a conditional vector is given to the output of a discriminator and the input of a generator.
	A combination of the C-CycleGAN and AC-GAN models has been proposed in the field of image generation \cite{naritomi2018foodchangelens, horita2018food, bozorgtabar2019using, choi2018stargan}.
	
	We focus on a composite model consisting of a CycleGAN model, which allows the domain transformation of unpaired data, and an AC-GAN model, which allows a discriminator to predict the class for locating a sound source inside a structure.
	Although the extended models of CycleGAN typically focus on generators, we focus on discriminators.
	The most common methods that focus on generators use conditional vectors as the input to a discriminator.
	However, we do not use conditional vectors as the input: our goal is to estimate the source location from real observations only; hence, conditional vectors cannot be used as inputs to the discriminator.
	Therefore, we propose a method that combines CycleGAN and AC-GAN models (Ac-CycleGAN) for locating a sound source inside a structure and demonstrate that this method outperforms the individual models.
	Furthermore,  we demonstrate that the Ac-CycleGAN discriminator categorizes the data  by utilizing only those portions characterized by minimal domain transformation errors.
	
	The remainder of this paper is organized as follows:
	In Section \ref{Sec:Def SSL inside structure}, we provide the definition and formulation of the problem of locating a sound source inside a structure.
	The proposed method is described in Section \ref{Sec:Proposed method}.
	Section \ref{Sec:Datasets} presents the simulation datasets and actual experimental datasets used in the study.
	Section \ref{Sec:Ac-CycleGAN} describes the implementation of Ac-CycleGAN.
	The results are described and discussed in Section \ref{Results and discussion}, and conclusions are presented in Section \ref{Sec:Conclusion}.
\section{Formulation of the problem of SSL inside a structure}\label{Sec:Def SSL inside structure}
	This section describes the formulation of the problem of locating a sound source inside a structure.

\subsection{Formulation}
	Consider a situation in which a noise source exists inside a structure, the structure is excited by the noise source, and the noise is observed outside the structure.
	The forced vibrations in an acoustic-structural coupled system that models the above physical phenomenon can be described using the finite discretization equation given below.
	\begin{equation}\label{Eq:FEM_1}
	\begin{split}
		\left\lbrack \begin{array}{cc}
		{\mathbf{M}}_{\mathrm{S}}  & 0\\
		\overline{\rho_0 } {\mathbf{R}}^{\mathrm{T}}  & {\mathbf{M}}_{\mathrm{F}}
		\end{array}\right\rbrack
		\left\lbrack \begin{array}{c}
		\ddot{\mathbf{u}} \\
		\ddot{\mathbf{p}} 
		\end{array}\right\rbrack +
		\left\lbrack \begin{array}{cc}
		{\mathbf{C}}_{\mathrm{S}}  & 0\\
		0 & {\mathbf{C}}_{\mathrm{F}} 
		\end{array}\right\rbrack 
		\left\lbrack \begin{array}{c}
		\dot{\mathbf{u}} \\
		\dot{\mathbf{p}} 
		\end{array}\right\rbrack + &
		\left\lbrack \begin{array}{cc}
		{\mathbf{K}}_{\mathrm{S}}  & -\mathbf{R}\\
		0 & {\mathbf{K}}_{\mathrm{F}} 
		\end{array}\right\rbrack \left\lbrack \begin{array}{c}
		\mathbf{u}\\
		\mathbf{p}
		\end{array}\right\rbrack 
		=
		\left\lbrack \begin{array}{c}
		0 \\
		{\mathbf{F}}_{\mathrm{F}}
		\end{array}\right\rbrack,
	\end{split}
	\end{equation}
	where  $\mathbf{u}$, $\mathbf{p}$, $\mathbf{M}$, $\mathbf{C}$, $\mathbf{F}$, and $\mathbf{K}$ represent the displacement, sound pressure, mass, damping, external force and stiffness matrices, respectively.
	Suffixes S and F indicate that the parameters correspond to structural and acoustic properties, respectively.
	Here, $\overline{\rho_0 }$ denotes the mass density constant of the acoustic fluid, and the matrix $\mathbf{R}$ denotes the coupled term.
	
	Locating a sound source inside a structure is an inverse problem of determining the input terms from observations on acceleration, velocity, or displacement under the condition that the mass, damping, stiffness, and coupling matrix are known.
	In general, Eq.(\ref{Eq:FEM_1}) has an infinite number of eigenvalues and is a singular point if decay terms are absent.
	The TDOA method attempts to estimate the location of a noise source based on multiple observation points at the resonance point; hence, TDOA cannot be used in this study because of the lack of phases.
	In other words, the solution is not guaranteed to be unique at the resonance point and the problem is ill-posed.
	In addition, the observed acoustic features are masked by the resonance characteristics derived from the acoustic space and structures, which hinders the determination of the source location.
	Therefore, the location of a sound source must be stochastically determined using machine learning methods and numerical analysis that facilitate the simulation of physical phenomena.
	
\subsection{Problem} 
\begin{figure*}[t!]
	\centering
	\includegraphics[width=140 mm]{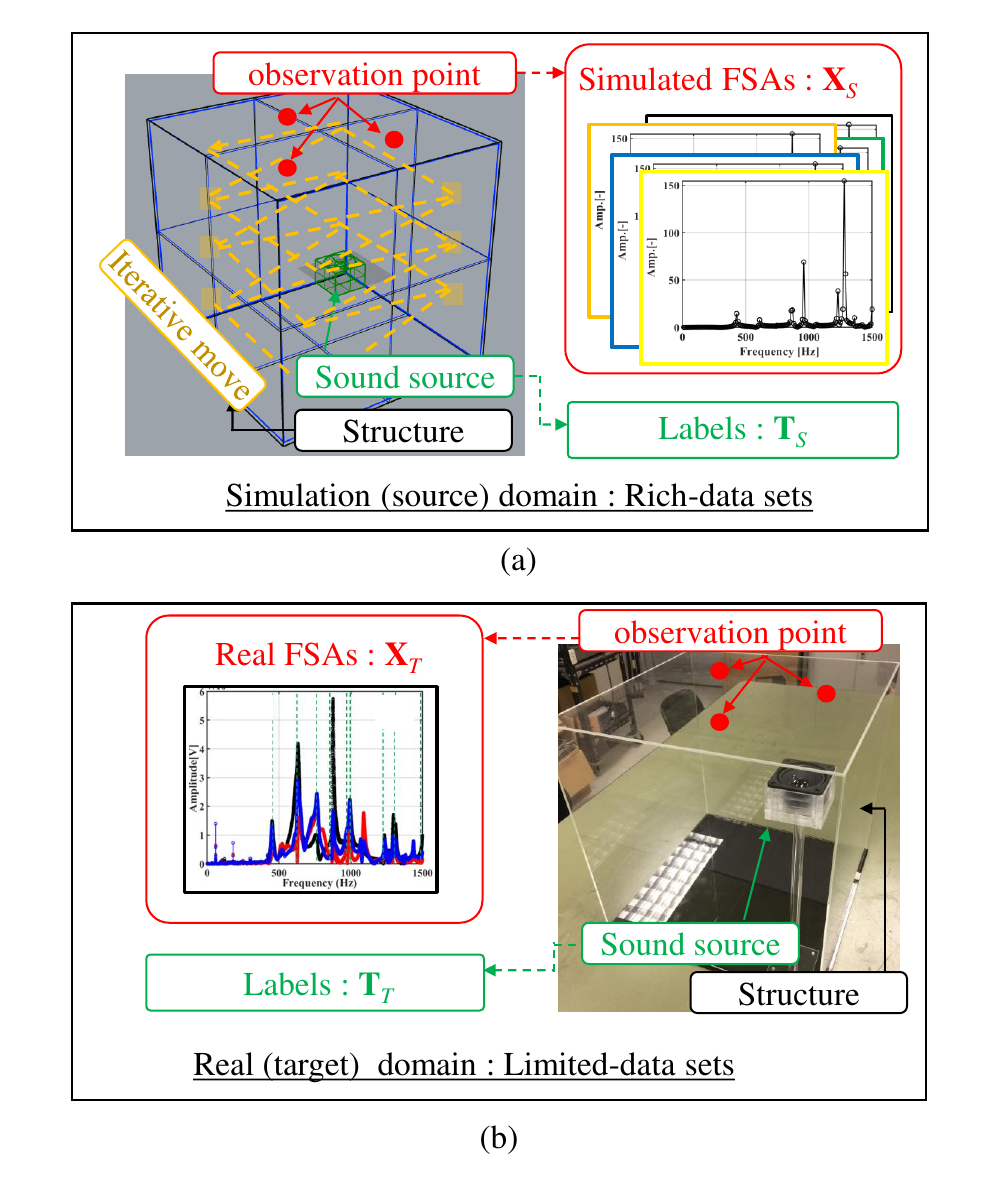}
	\caption{The problem of locating a sound source inside a structure. (a)The simulation domain. (b)The real domain.}
	\label{fig:Problem.pdf}
\end{figure*}

Consider a situation in which there is a noise source inside a structure.
The noise source causes the structure to vibrate.
The problem of locating the above noise source is formulated as follows: predict the location of the sound source using vibration data, which is measured using multiple acceleration sensors installed on the outer surface of the structure.
To solve this problem, SSL is conducted in a real environment using simulation datasets that can be generated numerically in large quantities and a small number of real datasets (Fig.~\ref{fig:Problem.pdf}).
The simulation and real domains correspond to source domain $S$ and target domain $T$, respectively.
The observed data obtained via the simulation is denoted by the D-dimensional input vector $\{{\mathbf{x}_{Si}}\}^N_{i=1} \in S$ and the real data is denoted by  $\{{\mathbf{x}_{Tj}}\}^M_{j=1} \in T$.
In addition, the locations of the noise source paired with the input data is defined as a K-dimensional label vector $\{{\mathbf{t}_{Si}}\}^N_{i=1} \in S$ and $\{{\mathbf{t}_{Tj}}\}^M_{j=1} \in T$.
The data matrix $\mathbf{X}$ and the label matrix $\mathbf{T}$ for each domain are given by
\begin{equation}\label{Eq:Input_X_data_S}
\mathbf{X}_S={\left({\mathbf{x}}_{S1} ,{\mathbf{x}}_{S2},{\mathbf{x}}_{S3} ,\cdots ,{\mathbf{x}}_{SN} \right)}^{\mathrm{T}},
\end{equation} 
\begin{equation}\label{Eq:Teach_data_S}
\mathbf{T}_S={\left({\mathbf{t}}_{S1} ,{\mathbf{t}}_{S2} ,{\mathbf{t}}_{S3} ,\cdots ,{\mathbf{t}}_{SN} \right)}^{\mathrm{T}},
\end{equation}
\begin{equation}\label{Eq:Input_X_data_T}
\mathbf{X}_T={\left({\mathbf{x}}_{T1} ,{\mathbf{x}}_{T2},{\mathbf{x}}_{T3} ,\cdots ,{\mathbf{x}}_{TM} \right)}^{\mathrm{T}},
\end{equation} 
\begin{equation}\label{Eq:Teach_data_T}
\mathbf{T}_T={\left({\mathbf{t}}_{T1} ,{\mathbf{t}}_{T2} ,{\mathbf{t}}_{T3} ,\cdots ,{\mathbf{t}}_{TM} \right)}^{\mathrm{T}}.
\end{equation}
In other words, the combination of the data expressed as ($\mathbf{X}, \mathbf{T}$) and observed outside the structure for each domain and the location of the noise source is treated as the dataset.
We assume that $N$ is approximately ten times larger than $M$, which satisfies the condition $M \ll  N$.

\section{Proposed method}\label{Sec:Proposed method}
	\begin{figure*}[!t] 
		\centering
		\includegraphics[width=140 mm]{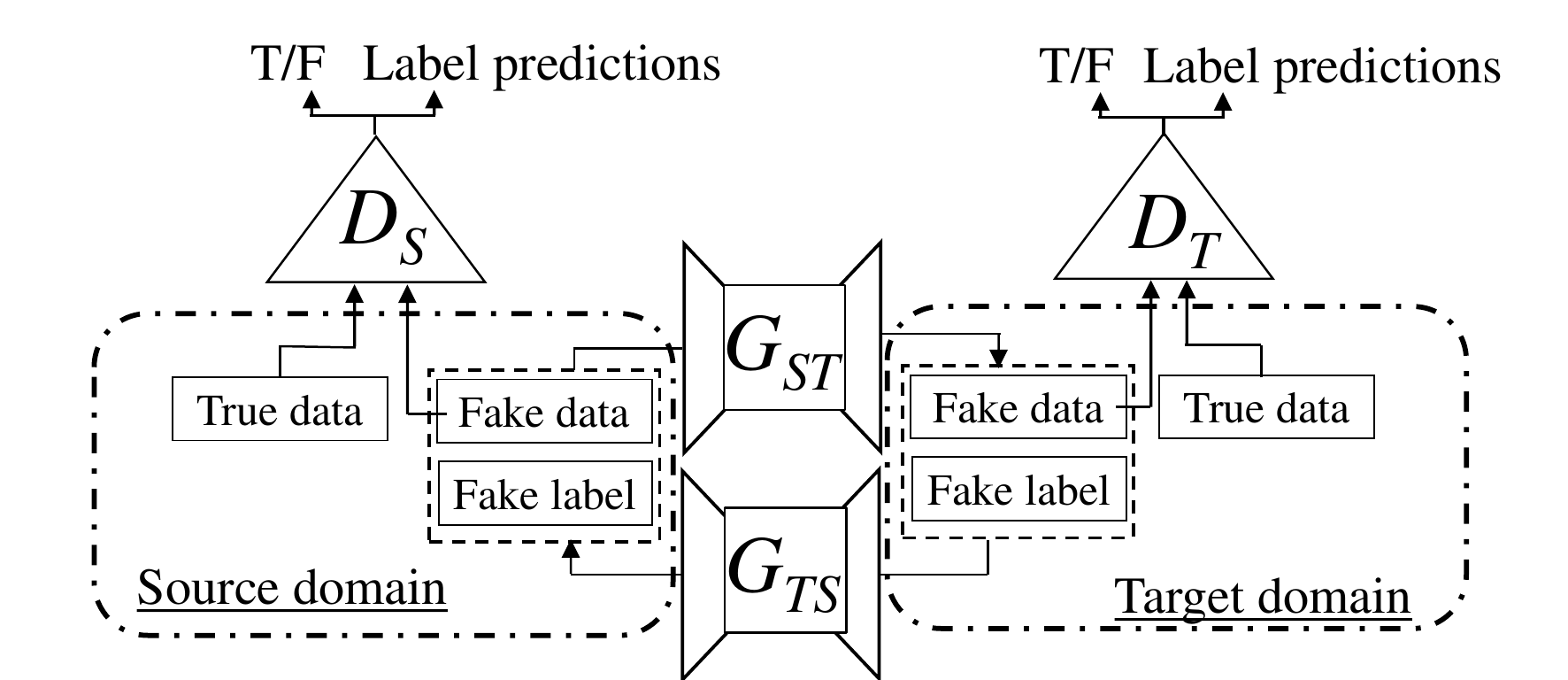}
		\caption{The Ac-CycleGAN. ${D}_S$ and ${D}_T$ represent the source and target domain discriminators, respectively. ${G}_{ST}$ and ${G}_{TS}$ represent the source to target and target to source generators, respectively.}
		\label{fig:Proposed_method.pdf}
	\end{figure*}
	Fig.~\ref{fig:Proposed_method.pdf} shows the Ac-CycleGAN model adopted in the study.
	In the figure, $D_S$ and $D_T$ denote the source and target domain discriminators, respectively.
	$G_{ST}$ and $G_{TS}$ represent the source-to-target and target-source generators, respectively.
	Through adversarial learning using $D_T$, $G_{ST}$ generates fake data that resemble the target domain.
	In contrast, through adversarial learning using $D_S$, $G_{TS}$ generates fake data that resemble the source domain.
	$D_T$ discriminates between the fake data generated by $G_{ST}$ and true target data, and $D_S$ discriminates between the fake data generated by $G_{TS}$ and true source data.
	
	The input and output of the generators are the fake data and labels.
	However, the input to the discriminators is only the data (true and fake) and not the labels.
	In addition, the discriminators output the true/fake decisions and label predictions.
	It is important not to use the conditional vector as the input to the discriminator, because the aim is to estimate the location of the sound source from only the observed data.
	The adversarial loss for mapping function $G_{ST}$ is formulated as follows.
	\begin{align}\label{Eq:LGAN}
	\mathcal{L}_{adv}(G_{ST}, D_T, \mathbf{x}_S, \mathbf{x}_T, \mathbf{t}_S) &= \mathbb{E}_{\mathbf{x}_T \sim p_{data}(\mathbf{x}_T)}\bigl[\log D_T(\mathbf{x}_T)\bigr] \notag \\
	&+\mathbb{E}_{(\mathbf{x}_S, \mathbf{t}_S)\sim p_{data}(\mathbf{x}_S, \mathbf{t}_S)}\biggl[\log \Bigl(1-D_T\bigl(G_{ST}(\mathbf{x}_S, \mathbf{t}_S)\bigr)\Bigr)\biggr].
	\end{align}
	$\mathcal{L}_{adv}(G_{TS}, D_S, \mathbf{x}_T, \mathbf{x}_S, \mathbf{t}_T)$ denotes the adversarial loss of the mapping function $G_{TS}$ with respect to the opposite direction.
	Eq. (\ref{Eq:LGAN}) is similar to that describing the adversarial loss in a CycleGAN model, except that negative log-likelihood is replaced by the MSE, as in an LSGAN (Eq. (\ref{Eq:Ladv})) \cite{mao2017least}.
	\begin{align}\label{Eq:Ladv}
	\mathcal{L}_{adv}(G_{ST}, D_T, \mathbf{x}_S, \mathbf{x}_T, \mathbf{t}_S) &= \mathbb{E}_{\mathbf{x}_T \sim p_{data}(\mathbf{x}_T)}\bigl[(D_T(\mathbf{x}_T)-1)^2\bigr] \notag \\
	&+\mathbb{E}_{(\mathbf{x}_S, \mathbf{t}_S)\sim p_{data}(\mathbf{x}_S, \mathbf{t}_S)}\biggl[D_T\bigl(G_{ST}(\mathbf{x}_S, \mathbf{t}_S)\bigr)^2\biggr].
	\end{align}
	The generator attempts to minimize the above objective, whereas the discriminator attempts to maximize it.
	Note that the labels are not input to the discriminator.
	This structure is similar to a part of a StarGAN model \cite{choi2018stargan}, which consists of multiple discriminators with an Ac-GAN structure and one generator.
	However, the discriminator in a StarGAN model is a domain classifier, and the information on the target domain is used in the generator.
	Therefore, the use of this method is not feasible in our study.
	
	The cycle consistency loss introduced by the two generators as well as the CycleGAN model, which incorporates the conditional vector, is given as follows.
	\begin{align}\label{Eq:LCYC}
	\mathcal{L}_{cyc}(G_{ST},G_{TS}) &= \mathbb{E}_{(\mathbf{x}_S, \mathbf{t}_S) \sim p_{data}(\mathbf{x}_S, \mathbf{t}_S)}\bigl[\|G_{TS}(G_{ST}(\mathbf{x}_S, \mathbf{t}_S))-(\mathbf{x}_S, \mathbf{t}_S)\|_1\bigr] \notag \\
	&+\mathbb{E}_{(\mathbf{x}_T, \mathbf{t}_T) \sim p_{data}(\mathbf{x}_T, \mathbf{t}_T)}\bigl[\|G_{ST}(G_{TS}(\mathbf{x}_T, \mathbf{t}_T))-(\mathbf{x}_T, \mathbf{t}_T)\|_1\bigr].
	\end{align}
	
	Furthermore, identity loss is employed because the model is based on the concatenation of the labels in the channel directions.
	The original paper reported that this loss function contributes to the domain identity for a channel \cite{zhu2017unpaired}.
	The identity loss formulated in this study is given as follows:
	\begin{align}\label{Eq:LIDE}
	\mathcal{L}_{identity}(G_{ST},G_{TS}) &= \mathbb{E}_{(\mathbf{x}_T, \mathbf{t}_T) \sim p_{data}(\mathbf{x}_T, \mathbf{t}_T)}\bigl[\|G_{ST}(\mathbf{x}_T, \mathbf{t}_T)-(\mathbf{x}_T, \mathbf{t}_T)\|_1\bigr] \notag \\
	&+\mathbb{E}_{(\mathbf{x}_S, \mathbf{t}_S) \sim p_{data}(\mathbf{x}_S, \mathbf{t}_S)}\bigl[\|G_{TS}(\mathbf{x}_S, \mathbf{t}_S)-(\mathbf{x}_S, \mathbf{t}_S)\|_1\bigr].
	\end{align}
	
	The auxiliary classifier loss introduced in the Ac-GAN model is applied to the discriminator to allow the discriminator to predict the sound source location.
	The auxiliary classifier loss introduced in the discriminator of the target domain is given as follows:
	\begin{align}\label{Eq:LC}
	L_C(G_{ST}, D_T, \mathbf{x}_S, \mathbf{t}_S, \mathbf{x}_T, \mathbf{t}_T) &= E_{(\mathbf{x}_T, \mathbf{t}_T) \sim p_{data}(\mathbf{x}_T, \mathbf{t}_T)}[\log D_T(T=\mathbf{t}_T|\mathbf{x}_T)] \notag \\
	&+ E_{(\mathbf{x}_S, \mathbf{t}_S) \sim p_{data}(\mathbf{x}_S, \mathbf{t}_S)}[\log D_T(T=\mathbf{t}_T|G_{ST}(\mathbf{x}_T))].
	\end{align}
	The auxiliary classifier loss in the source-domain discriminator is $L_C(G_{TS}, D_S, \mathbf{x}_T, \mathbf{t}_T, \mathbf{x}_S, \mathbf{t}_S)$.

	Finally, the objective function to be optimized is expressed as 
	\begin{align}\label{Eq:LFULL}
	\mathcal{L}(G_{ST}, G_{TS}, D_T, D_S)&=\mathcal{L}_{adv}(G_{ST}, D_T, \mathbf{x}_S, \mathbf{x}_T, \mathbf{t}_S) \notag \\
	&+ \mathcal{L}_{adv}(G_{TS}, D_S, \mathbf{x}_T, \mathbf{x}_S, \mathbf{t}_T) \notag \\
	&+ \lambda_{cyc}\mathcal{L}_{cyc}(G_{ST},G_{TS}) \notag \\
	&+ \lambda_{identity}\mathcal{L}_{identity}(G_{ST},G_{TS}) \notag \\
	&\pm L_C(G_{ST}, D_T, \mathbf{x}_S, \mathbf{t}_S, \mathbf{x}_T, \mathbf{t}_T) \notag \\
	&\pm L_C(G_{TS}, D_S, \mathbf{x}_T, \mathbf{t}_T, \mathbf{x}_S, \mathbf{t}_S),
	\end{align}
	where $\lambda_{cyc}$ and $\lambda_{identity}$ are the hyperparameters that control the relative importance of domain reconstruction and identity losses, respectively.
	The Ac-CycleGAN model is trained on the basis of the following equation:
	\begin{align}\label{Eq:LFULL_minmax}
	D_S^*, D_T^* = \arg \min_{G_{ST}, G_{TS}}\max_{D_S,D_T} \mathcal{L}(G_{ST}, G_{TS}, D_T, D_S)
	\end{align}
	Note that in both the generator and discriminator, the plus or minus sign of $L_C$ is chosen according to the direction of minimization.
	
\section{Datasets of the simulation and the actual experimental data}\label{Sec:Datasets}
	\begin{figure}[t] 
  \centering
  \includegraphics[width=110 mm]{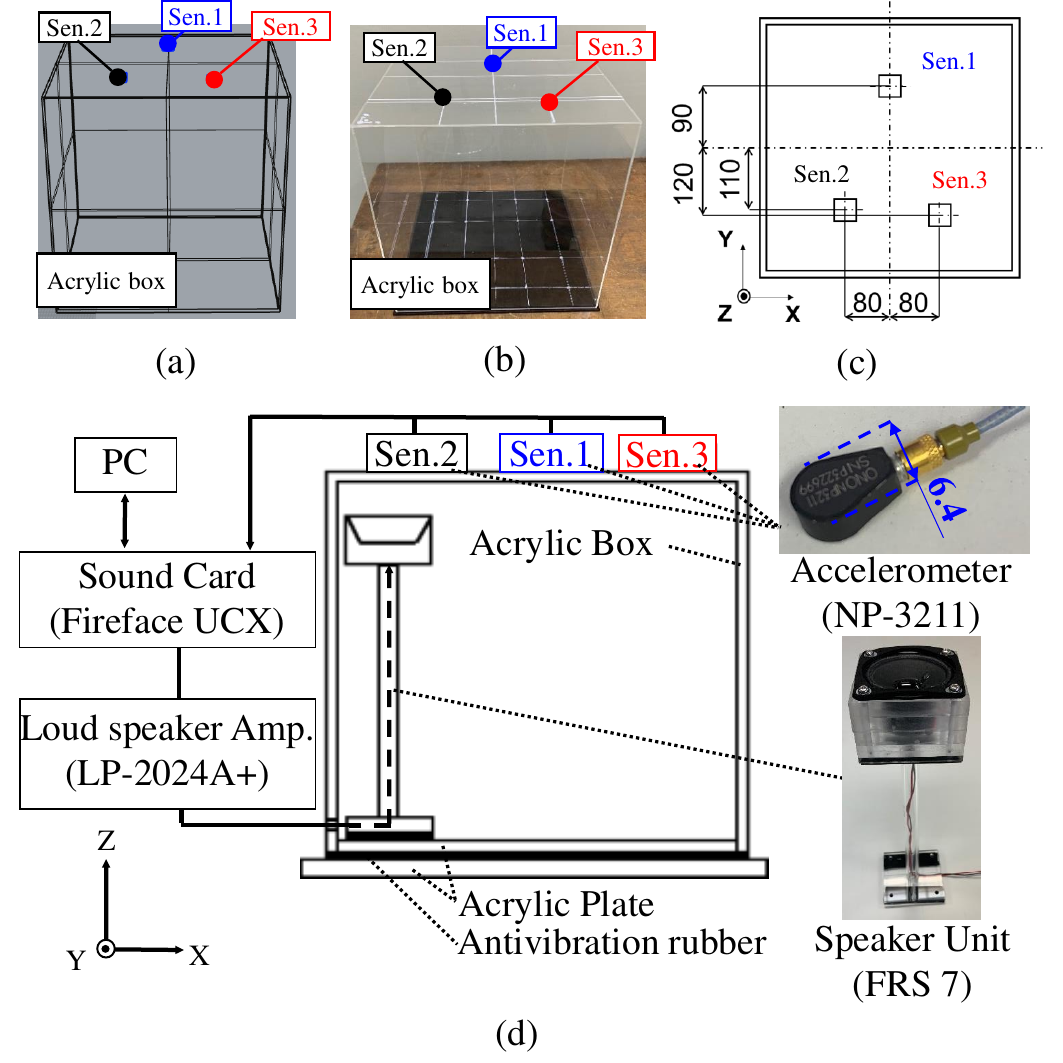}
  \caption{Simulation and real domain setup. (a) Simulation. (b) Real. (c) Sensor placement positions. Units in the figure are in (mm). (d) Actual experimental setup. The dataset are the same as that used in our previous study \cite{kita2021fundamental}.}
  \label{fig:Datasets.pdf}
	\end{figure}
	The datasets used are the same as those used in our previous study \cite{kita2021fundamental}.
	It is assumed that the acoustic excitation from a single sound source within the structure (acrylic box) is measured using three accelerometers mounted on the outer surface of the structure.
	The frequency spectra of the accelerometer (FSA) are used as the observation data.
	Fig.~\ref{fig:Datasets.pdf} shows the simulation domain, real domain, and experimental setup used for sampling the data in the real domain.
	The size of the structure and the location of the accelerometers are the same in the simulation and real domains.
	\begin{table}[t] 
		\caption{Analysis conditions}
		\vspace{-1mm}
		\label{Tab:Analysis conditions}
		\begin{center}
			\begin{tabular}{l r} 
				\bhline{1pt}
				Acrylic young's modulus & 27 MPa \\
				Acrylic density & 1180 \si{kg/m}$^3$ \\
				Acrylic damping ratio & 0.8 \\
				Interval between movements of the sound source  & 50 \si{mm} \\
				Acoustic volume & 400 $\times$ 400 $\times$ 400 \si{mm}$^3$ \\
				Thickness of the acrylic box & 3 \si{mm} \\
				Number of data points & 512 \\
				Observation & Sen.1 -- Sen.3 \\
				Frequency range & 0.01--1.5 \si{kHz} \\
				\bhline{1pt}
			\end{tabular}
		\end{center}
	\end{table}
	The simulation conditions are presented in Table~\ref{Tab:Analysis conditions}.
	The simulation data are generated from a coupled acoustic structure analysis using the finite element method (FEM).
	The FEM solver is a full harmonic analysis in ANSYS Mechanical \cite{Ansys}.
	The frequency range is 0.01--1.5 kHz, and the increment range is 10 Hz.
	The actual experimental conditions are shown in Fig.~\ref{fig:Datasets.pdf} (d).
	The noise is generated using a PC and emitted from a loudspeaker (Visation FRS 7) via a sound card (Fireface UCX) and a loudspeaker amplifier (LP-2024-A +).
	The acoustic excitation of the structure is measured using three acceleration sensors (Ono Sokki Co., Ltd. NP-3211) that are installed on the outer surface of the structure.
	\begin{table}[t]
		\caption{Measurement conditions}
		\vspace{-1mm}
		\label{Tab:Measurement conditions}
		\begin{center}
			\begin{tabular}{l r} 
				\bhline{1pt}
				Interval between movements of the sound source &  100 \si{mm} \\
				Number of data points & 64 \\
				Observation & Sen.1 -- Sen.3 \\
				Input signal & Swept sinusoidal \\
				Frequency range & 0.01--1.5 \si{kHz} \\
				Sampling rate & 4.8 \si{kHz} \\
				Sound pressure & 90 \si{dB} at 1 \si{m} \\
				Subband width & 10 \si{Hz} \\
				\bhline{1pt}
			\end{tabular}
		\end{center}
	\end{table}
	The experimental conditions are presented in Table~\ref{Tab:Measurement conditions}.
	The main difference between the simulation and real environments is that the number of data points in the former is 512, whereas it is 64 in the latter.
	A frequency range of 0.01--1.5 kHz and a 150-dimensional vector per sensor input are set for both the simulated and real environments.
	The time-series acceleration data measured using the three sensors are transformed into FSA by applying fast Fourier transform.

	The horizontal concatenation and transposition of the FSAs observed from the three accelerometers are treated using the Ac-CycleGAN model.
	That is, the size of the observation data obtained from each source point is 150 $\times$ 3 ($\mathbf{x}_{Si}, \mathbf{x}_{Tj} \in \mathbb{R}^{150 \times 3}$).
	Therefore, the data are treated as image data \cite{70216d058af64a5f98c208ef90894204, 6d68305e44514951b940a0223d30a0af}.
	In this case, to test the performance of the Ac-CycleGAN model in solving  the classification problem, the labels are defined by dividing the internal acoustic space into eight equal parts, which are then labeled using one of the K coding schemes ($\mathbf{t}_{Si}, \mathbf{t}_{Tj} \in \mathbb{R}^{8}$).
	The percentage of actual experimental data used as training data for the Ac-CycleGAN model varies from $20$ to $80\%$.
	
\section{Implementation of the Ac-CycleGAN}\label{Sec:Ac-CycleGAN}
	\begin{table}[t]
	\caption{Ac-CycleGAN Generator.}
	\vspace{-1mm}
	\label{Tab:Ac-cycleGAN Generator}
	\begin{center}
	\begin{tabular}{wl{15mm}r}
	\bhline{1pt}
	Layer&
	\begin{tabular}{r}
	Input (Label) : EM(8, 50), FC450 \\
	Reshape to (150, 3, 8)\\
	Input (Image) : (150, 3, 1) \\
  Concatenate : (150, 3, 9) \\
  Conv-IN-LeakyReLU : FI64, FI128, FI256, FI512 \\
  Res block : C512 $\times$ 9\\
  ConvT-IN-LeakyReLU : FI256, FI128, FI64 \\
  Conv-IN : FI1 \\
	Output (Label) : Flatten, FC8\\
	Output (Image) : Output to activation\\
	\end{tabular}\\
	Output activation&
	\begin{tabular}{r}
	Label:softmax\\
	Image:tanh\\
	\end{tabular}\\
	Optimization&
	\begin{tabular}{r}
	Adam : Initial learning rate = $0.0002$\\
	($\beta_{1}$ = 0.5, $\beta_{2}$ = 0.999) \\
	\end{tabular}\\
	Loss function & Image : Mean squared error, Label : Cross entropy error \\
	Initialization & RandomNormal (standard deviation : 0.02)\\
	\bhline{1pt}
	\end{tabular}
	\end{center}
	\end{table}
	
	\begin{table}[t]
	\caption{Ac-CycleGAN Discriminator.}
	\vspace{-1mm}
	\label{Tab:Ac-cycleGAN Discriminator}
	\begin{center}
	\begin{tabular}{wl{15mm}r}
	\bhline{1pt}
	Layer&
	\begin{tabular}{r}
	Input (Image) : (150, 3, 1)\\
	Conv-LeakyReLU : FI64\\
	Conv-IN-LeakyReLU : FI128, FI256, FI512, FI512\\
	FI1 (Patch out)\\
	Output (Label) : Flatten, FC8\\
	Output (Image) : Output to activation
	\end{tabular}\\
	Output activation&
	\begin{tabular}{r}
	Label:softmax \\
	Output:-\\
	\end{tabular}\\
	Optimization&
	\begin{tabular}{r}
	Adam : Learning rate = $0.0002$\\
	($\beta_{1}$ = 0.5, $\beta_{2}$ = 0.999) \\
	\end{tabular}\\
	Loss function & Image : Mean squared error, Label : Cross entropy error \\
	Initialization & RandomNormal (standard deviation : 0.02)\\
	\bhline{1pt}
	\end{tabular}
	\end{center}
	\end{table}
	The generator and discriminator in the Ac-CycleGAN model are a residual net (Resnet) \cite{he2016deep} and a patchGAN \cite{isola2017image}, respectively, and their respective settings are listed in Tables~\ref{Tab:Ac-cycleGAN Generator} and \ref{Tab:Ac-cycleGAN Discriminator}.
	In the preprocessing stage, the image is changed in the range [-1, 1].
	The generator has an encoder--decoder architecture.
	The model takes the source image and labels and generates the target image and label.
	This network contains several convolution units (Conv), residual blocks, and transposed convolution units (ConvT) \cite{zeiler2010deconvolutional}.
	The encoder and decoder consist of a Conv, instance normalization (IN) layer\cite{ulyanov2016instance}, and LeakyReLU \cite{maas2013rectifier}.
	The label is embedded in an 8 $\times$ 50 vector using an embedding layer (EM).
	Its output is passed through a fully connected layer (FC), reshaped to the same size as the image, and concatenated with the image in the channel direction.
	On the output side, the label is flattened and the number of class units is determined by the FC.
	For the discriminator, the receptive field of the image data for one output pixel is an image patch of 6 × 3 pixels.
	In the generator and discriminator, the 2-D convolution layers are set with a kernel size of (2, 3) and stride of (1, 1). The same padding is used for all Conv layers and most of the number of filters (FI) are the same as that used in the original study.
	Data augmentation with masking is used to construct the Ac-CycleGAN model \cite{zhong2020random, devries2017improved}.
	In addition, similarly to the CycleGAN implementation, the number of updates for the discriminator is half of that for the generator.
	The batch size and number of epochs are 1 and 200, respectively.
	The learning rate of the discriminator decreases linearly to zero after 100 epochs.
	$\lambda_{cyc}$ and $\lambda_{identity}$ are 10 and 5, respectively.
	The hold-out validation strategy is used to evaluate the model and the performance is tested for every percentage of the semi-supervised data.
	Accuracy (Acc.), as given by Eq.~ (\ref{Eq:Acc}), is used to evaluate the performance of the classifier model.
	\begin{equation}\label{Eq:Acc}
	\mathrm{Acc}\ldotp =\frac{\mathrm{The}\;\mathrm{number}\;\mathrm{of}\;\mathrm{correct}\;\mathrm{answers}}{\mathrm{The}\;\mathrm{number}\;\mathrm{of}\;\mathrm{test}\;\mathrm{data}}.
	\end{equation}

\section{Results and discussion}\label{Results and discussion}
	First, the performance of the AC-CycleGAN in SSL is described, followed by a comparison of the performance with that of a conventional method.
	Next, we explain the reasons why the discriminator is able to predict the location of a sound source.
	The output of the layer with mixed labels and image features is visualized in three dimensions using t-SNE \cite{van2008visualizing} to demonstrate that the features are separated based on their classes.
	Furthermore, we use gradient-weighted class activation mapping (Grad-CAM) \cite{selvaraju2017grad} to understand the reason behind this separation.
	The Grad-CAM results show that the discriminator focuses on the features necessary for label estimation and classification, even if the generator does not perfectly transform the domains.

\subsection{Performance with the Ac-CycleGAN}
	\begin{figure*}[t] 
		\centering
		\includegraphics[width=90 mm]{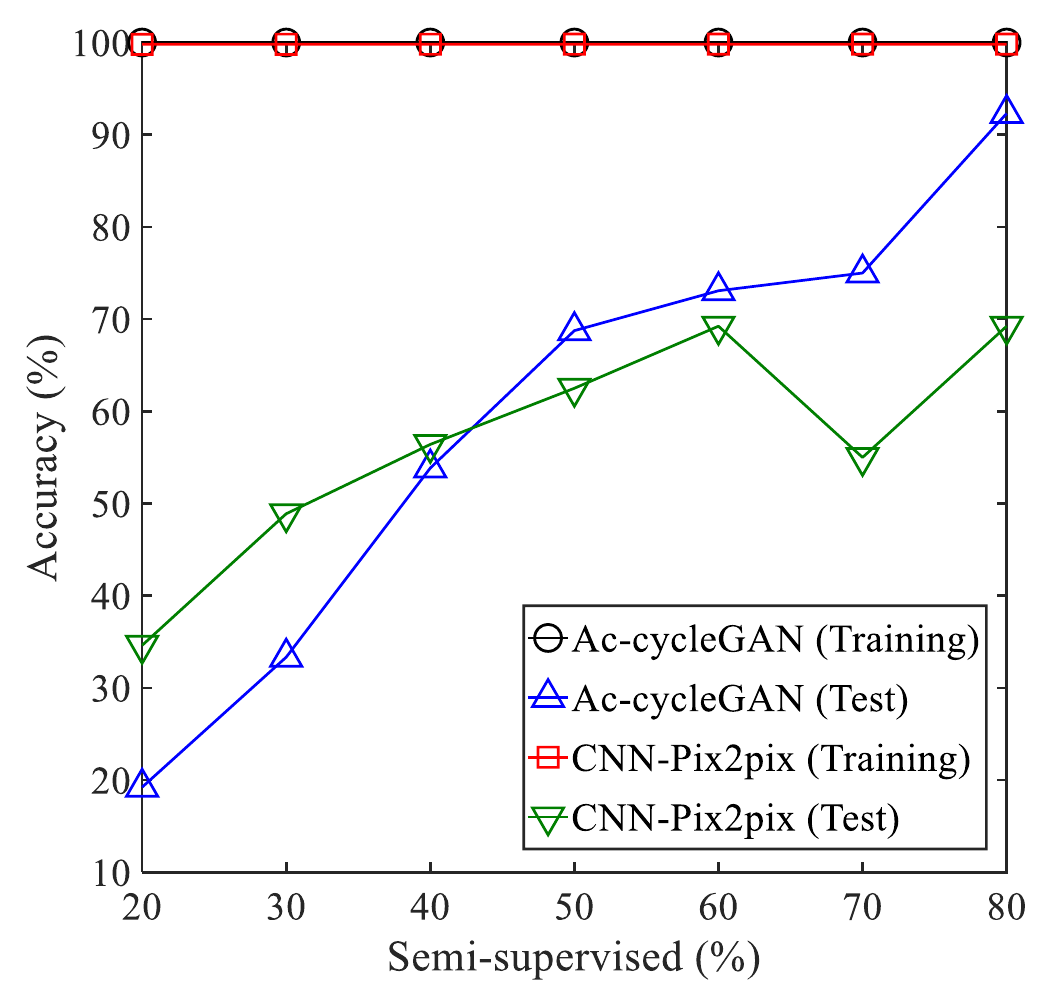}
		\caption{Results of classification. The horizontal axis shows the percentage of the total real environment data. As this percentage increases, the SSL problem becomes easier to solve as the conditions become closer to supervised condition.}
		\label{fig:Class.pdf}
	\end{figure*}
	\begin{table}[t]
		\caption{Proposed method (Ac-CycleGAN) vs. non-adaptation and conventional methods (CNN-Pix2pix).}
		\label{table:Sim to real.}
		\centering
		\begin{tabular}{wc{25mm}|wc{30mm}|wc{15mm}|wc{12mm}}\bhline{1pt}
			\multirow{1}{*}{Prob. (Criteria)}&\multirow{1}{*}{Model or method}&\multicolumn{1}{c|}{Training}&\multicolumn{1}{c}{Test}\\
			\hline
			\multirow{3}{*}{Class. (Acc.)}&\multirow{1}{*}{Non-adaptation}&99.75\%&15.38\%\\ \cline{2-4}
			&\multirow{1}{*}{CNN-Pix2pix}&99.82\%&69.23\%\\ \cline{2-4}
			&\multirow{1}{*}{Ac-CycleGAN}&100\%&92.31\%\\ \cline{1-1} \cline{2-4}
			\bhline{1pt}
		\end{tabular}
	\end{table}
	Fig.~\ref{fig:Class.pdf} shows the performance of the SSL models with respect to the ratio of semi-supervised data.
	``CNN-Pix2pix" denotes the conventional method and represents a model in which the image data is transformed from the target domain to the source domain using the DT model (pix2pix) and classified by a CNN constructed in the source domain.
	Both the Ac-CycleGAN and ``CNN-Pix2pix" exhibit approximately 100\% accuracy for the training data.
	For the test data, the conventional SSL method using the DT model is more accurate for up to 40\% of the semi-supervised data, whereas the proposed method is more accurate when the ratio of semi-supervised data is more than 50\%.

	Table \ref{table:Sim to real.} presents the performance of the proposed, conventional, and non-adaptation methods.
	These results are obtained at a semi-supervised data ratio of 80\%.
	Evidently, the proposed method outperforms the other methods, indicating that the performance in SSL is improved compared to that exhibited by the non-adaptation method.
	In our previous study \cite{kita2021fundamental}, we were unable to build models directly in a real domain because of small amounts of available data.

\subsection{Discriminator analysis}
	\begin{figure}[t]
		\centering
		\includegraphics[width=75 mm]{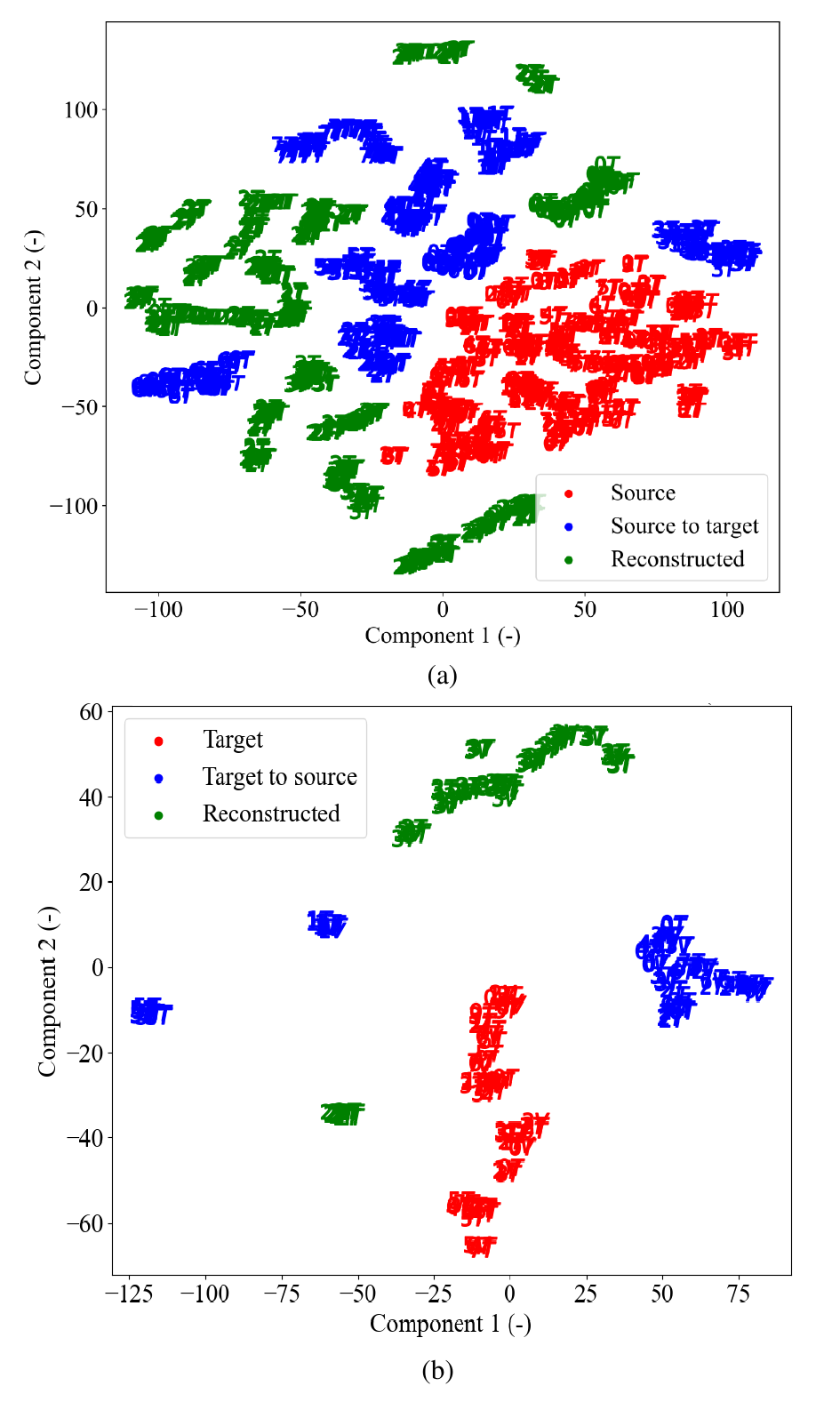}
		\caption{Results of generator reconstruction. (a) Source domain. (b) Target domain.}
		\label{fig:Reconstruct.pdf}
	\end{figure}
	To understand in more detail the functionality of the Ac-CycleGAN, let us first visualize the distribution of the data generated by the generator.
	Fig.~\ref{fig:Reconstruct.pdf} (a) and (b) show the results of t-SNE visualization for the distribution of the original (red), transformed (blue), and reconstructed (green) data in the source and target domains, respectively.
	The numbers in the figures represent classes, and ``T" and ``V" represent the training and the test data.
	The percentage of semi-supervised data is 80\%.
	The hyperparameters used for t-SNE visualization are perplexity 5, learning rate 5, and the number of iteration 5000.
	The discrepancy in the distribution indicates that the image reconstruction has failed.
	\begin{figure*}[t]
		\centering
		\includegraphics[width=75 mm]{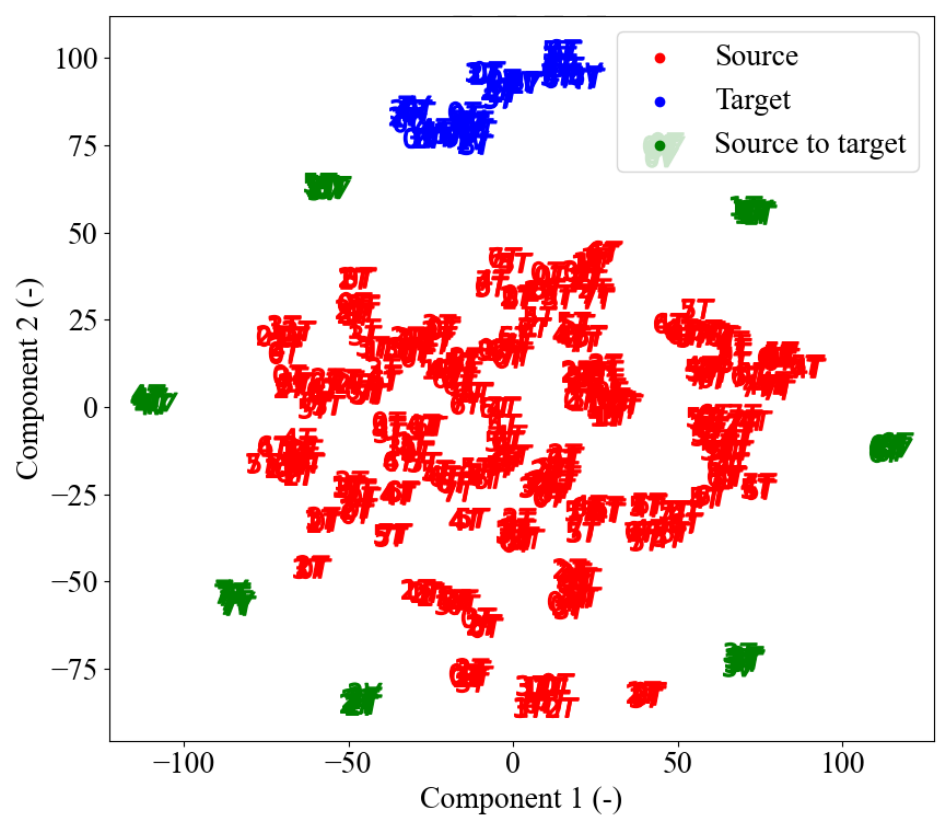}
		\caption{Results of generator transformation. Source to target.}
		\label{fig:Transfer.pdf}
	\end{figure*}
	Fig.~\ref{fig:Transfer.pdf} shows the generator's performance in transforming the source data into target data.
	Red, blue, and green represent source, target, and transformed data, respectively.
	As can be observed from the figure, the transformation from the source domain to the target domain is unsuccessful.
	In our previous study \cite{kita2023}, target-to-source transformation using pix2pix was successful; however, reconstruction and transformation using Ac-CycleGAN failed.
	The transformation fails may be because real data are complex and include nonlinearities, and recreating real data from simulated data is difficult.
	Although the transformation is inefficient, the discriminator exhibits an excellent performance in SSL.
	\begin{figure*}[t]
		\centering
		\includegraphics[width=150 mm]{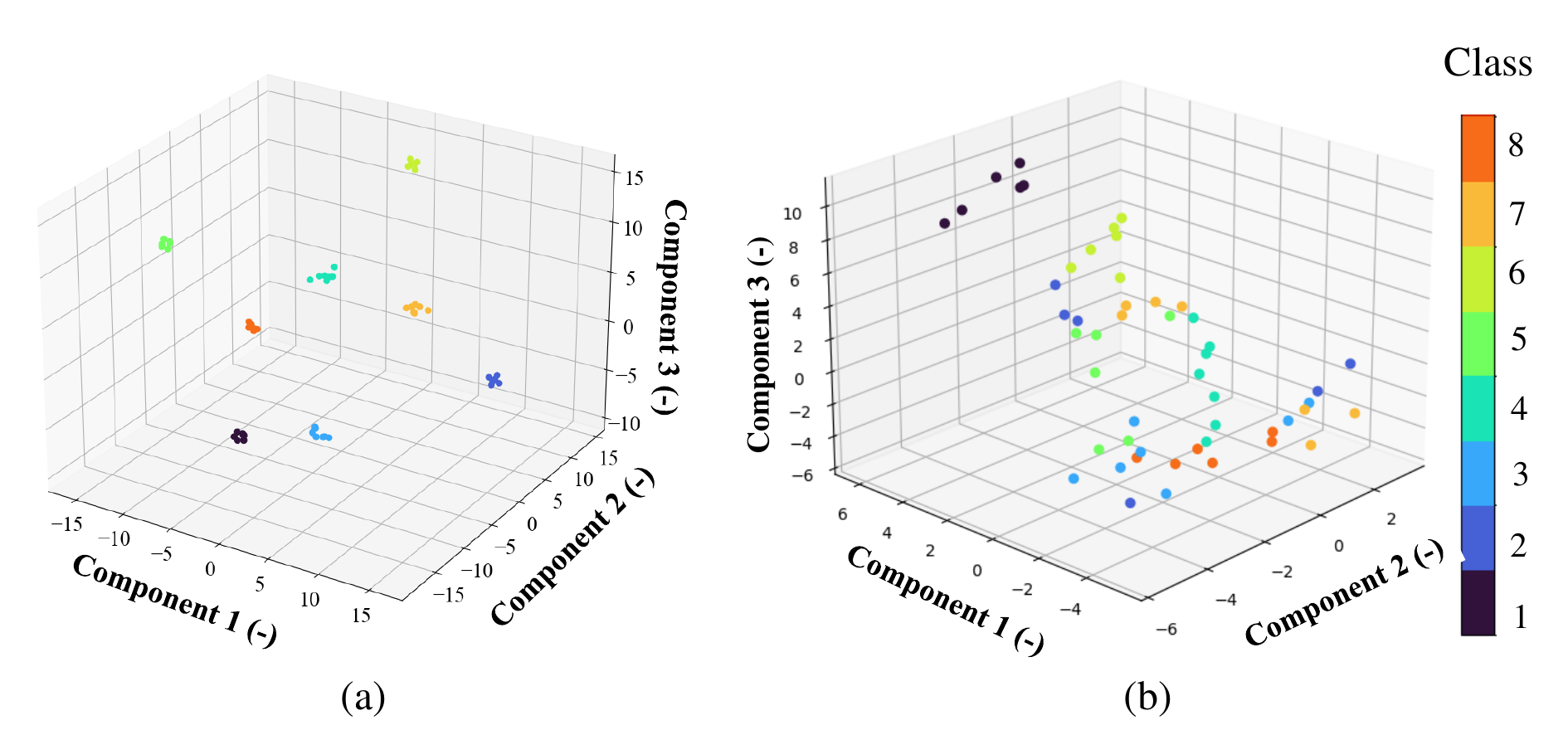}
		\vspace{-1mm}
		\caption{The output of the discriminator's hidden layer is visualized in three dimensions via t-SNE. (a) The output of the discriminator for the fake data generated by the generator. (b) The output of discriminator for the true data.}
		\label{fig:D_tsne.pdf}
	\end{figure*}
	\begin{figure*}[t]
		\centering
		\includegraphics[width=140 mm]{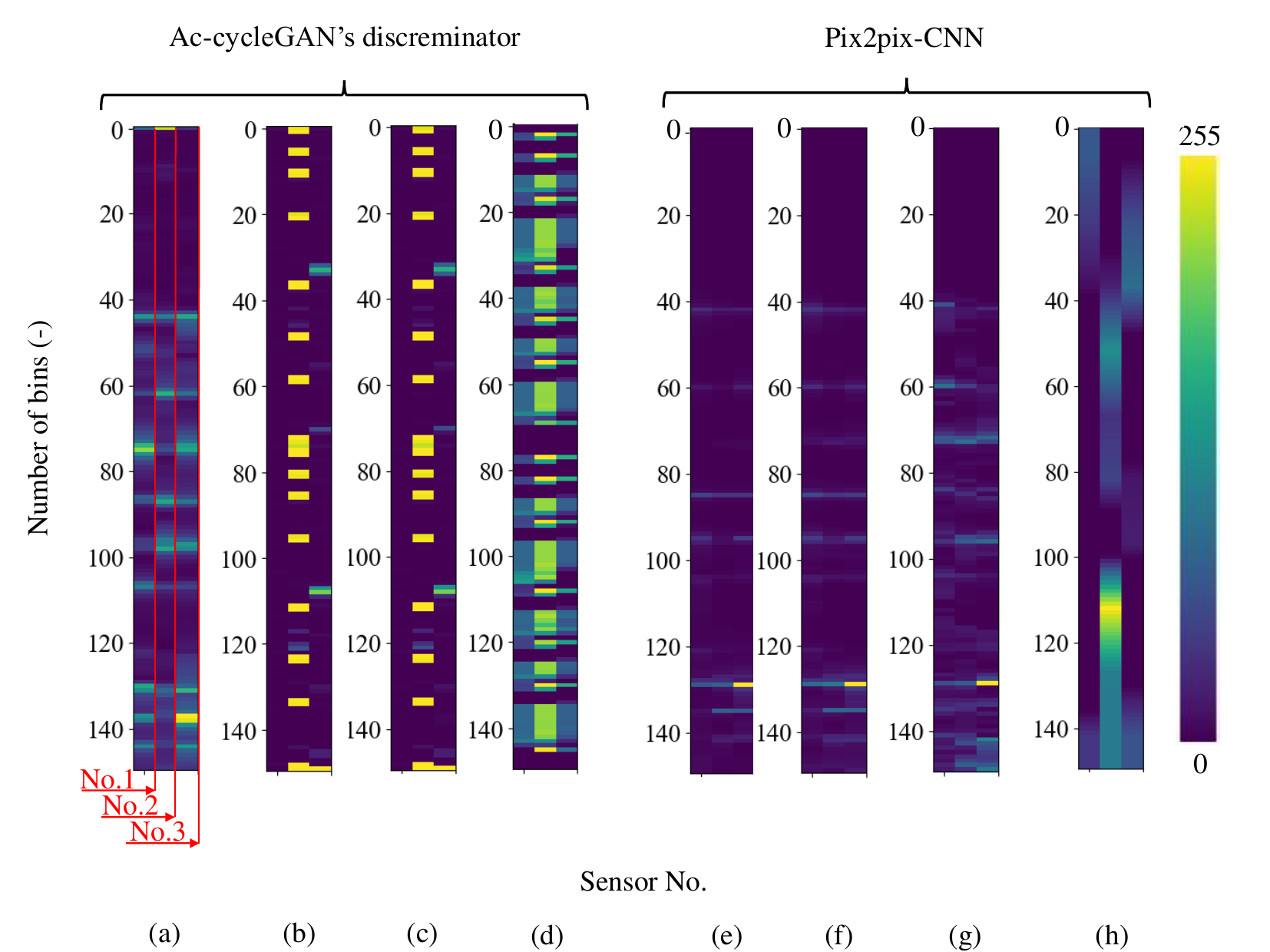}
		\caption{Discriminator's attention to the transformed data. (a)--(d) are for the Ac-CycleGAN discriminator. (a) Source data. (b) Transformed data. (c) The difference between the source and transformed data. (d) Grad-CAM results for the Ac-CycleGAN discriminator. (e)--(h) represent conventional methods that use a convolutional neural network (CNN) as a discriminator. (e) Target data. (f) Transformed data. (g) The difference between the target and transformed data. (h) Grad-CAM of CNN.}
		\label{fig:gradCAM.pdf}
	\end{figure*}
	\vspace{-0.1mm}
	
	We focus on the Ac-CycleGAN discriminator to understand which parts of the image data are used for the classification.
	Fig.~\ref{fig:D_tsne.pdf} shows a 3D t-SNE plot of the output from the hidden layer of the discriminator.
	The target hidden layer is the hidden layer before the last layer, which contains the mixture of ``patch out" for the true/fake decision and the label features.
	Fig.~\ref{fig:D_tsne.pdf} (a) and (b) show the output of the hidden layer when fake and true data are input to the discriminator, respectively.
	These figures show that the discriminator is able to classify according to labels, completely for the fake data generated by the generator and mostly for the true data.
	Furthermore, we focus on this hidden layer. Grad-CAM is used to visualize parts of the generated data that are important for the discriminator to classify.
	Fig.~\ref{fig:gradCAM.pdf} (a)--(d) show the source data, transformed data, difference between the source and transformed data, and the image visualized via Grad-CAM.
	The images are of size 150 $\times$ 3, and the pixel values of each image are normalized in the range of 0 to 255.
	Each image row contains sensor numbers 1 -- 3, from left to right.
	In particular, Fig.~\ref{fig:gradCAM.pdf} (b) shows that the generator transformation is imperfect: the results show large output values in some areas.
	A comparison between Fig.~\ref{fig:gradCAM.pdf} (c) (transformation errors) and Fig.~\ref{fig:gradCAM.pdf} (d) (Grad-CAM) reveals that the regions corresponding to large errors do not contribute to class discrimination and that the discriminator focuses on regions corresponding to small transformation errors for classification.
	Fig.~\ref{fig:gradCAM.pdf} (e)--(h) show the components of a conventional discriminator.
	These figures (e)--(h) show the target data, transformed data, difference between the target and transformed data, and images visualized by Grad-CAM, respectively.
	Note that (e) is the target data because the conventional method uses the ``target-to-source” transformation.
	These figures show that the generator efficiently performs the transformation; however, the discriminator uses the entire image as a feature for classification.
	This indicates that, the generator transformation does not necessarily need to be perfect for training the discriminator.
	\begin{figure}[t]
		\centering
		\includegraphics[width=100 mm]{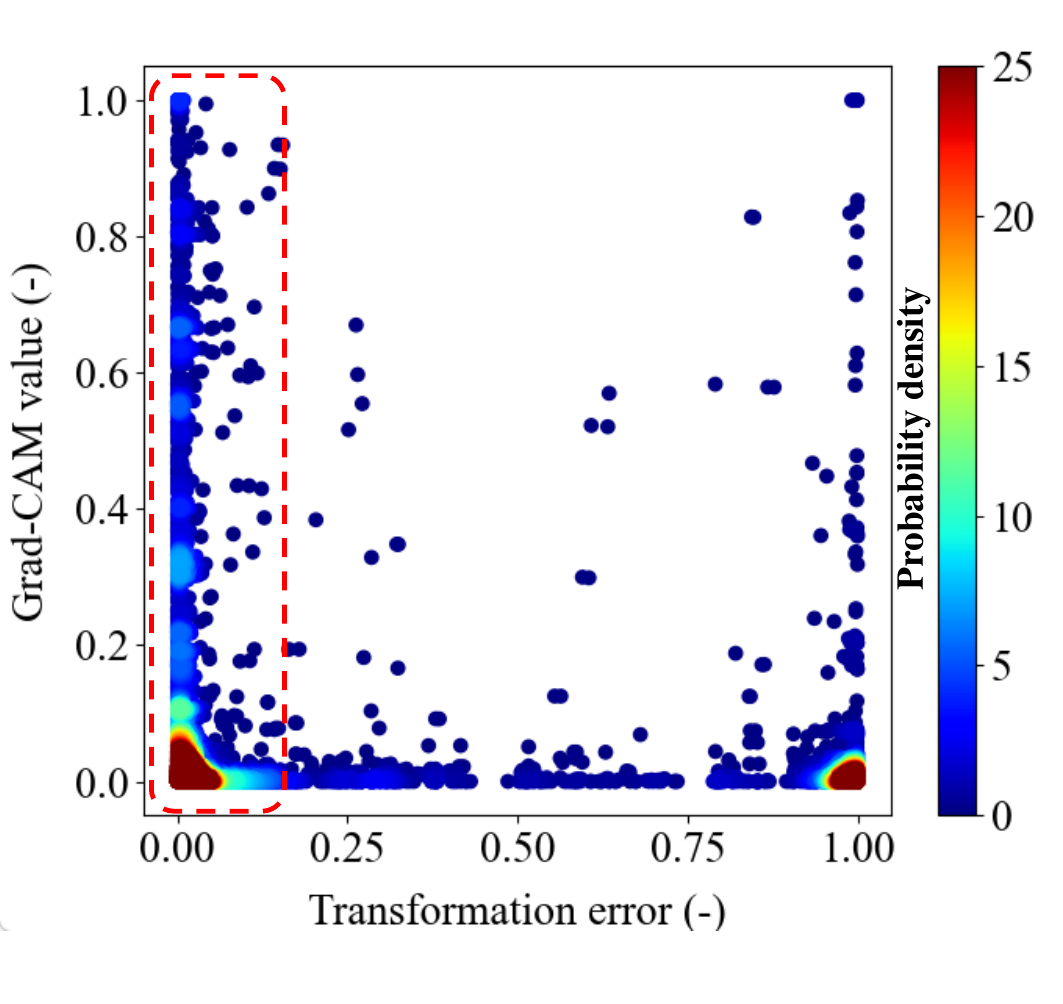}
		\vspace{-1mm}
		\caption{Relationship between transformation errors and Grad-CAM values.}
		\label{fig:kde.pdf}
	\end{figure}
	Since Fig.~\ref{fig:gradCAM.pdf} shows the visualization for a single image, all transformation data are statistically evaluated to investigate whether the transformation error and Grad-CAM value are inversely related.
	Fig.~\ref{fig:kde.pdf} shows the relationship between the normalized Grad-CAM value and the transformation error.
	The color contours represent the probability density obtained using the kernel density estimation method \cite{parzen1962estimation}.
	This figure shows that for large values of the transformation error, the density is clustered in areas where the Grad-CAM value is small.
	Furthermore, the density is clustered in areas where the transformation error and Grad-CAM value are small; however, the Grad-CAM value is spread over the entire area where the transformation error is small.
	This indicates that the Ac-CycleGAN's discriminator classifies labels by focusing on image areas that produce small transformation errors (red dotted-line in Fig.~\ref{fig:kde.pdf}).
	The concentration of density in regions corresponding to small values of these two variables may be due to the other convolution layers that extract the features.
	However, additional examination is difficult because this discriminator has many convolutional layers, and Grad-CAM is applied to the last or previous convolutional layer.

\section{Conclusions}\label{Sec:Conclusion}
We propose Ac-CycleGAN, which integrates an Ac-GAN and a CycleGAN, to solve the problem of predicting the location of a sound source inside a structure.
The proposed model facilitates the prediction of the labels for unpaired data between domains.
The proposed method was evaluated using the FSA data observed on the outer surface of the structure obtained from the simulation and real environments, and sound source locations as labels.
This method outperforms conventional methods that used the SSL and DT models.
The accuracy of the conventional method for SSL inside a structure is 69.23\%, whereas that of the proposed method is 92.31\%.
The two-dimensional distribution of t-SNE indicates that the data generated by the generator are not suitable for the reconstruction and transformation of the source domain into the target domain.
However, the three-dimension t-SNE visualization shows that the true and fake data produced by the generator are classified by the discriminator in the mixed layer of the image and label features.
Furthermore, Grad-CAM indicates that the discriminator does not capture and classify the features of the entire transformed data but only those parts of the generator that exhibit low transformation errors.

In this study, we focused on the classification problem to evaluate the discriminator performance in detail; however, in the future, we will focus on the regression problem.
To solve this problem, a method for embedding continuous values in layers is required, and we will investigate a new model for this purpose.
In addition, we plan to address the sensor placement problem to evaluate the robustness of the sound source features.
Furthermore, we plan to develop an SSL method that operates under unsupervised conditions.

	\section{Acknowledgements} 
	This study is financially supported by the JSPS KAKENHI (22K03991).
	\bibliography{Manuscript}

\begin{thebibliography}{52}
\expandafter\ifx\csname natexlab\endcsname\relax\def\natexlab#1{#1}\fi
\providecommand{\url}[1]{\texttt{#1}}
\providecommand{\href}[2]{#2}
\providecommand{\path}[1]{#1}
\providecommand{\DOIprefix}{doi:}
\providecommand{\ArXivprefix}{arXiv:}
\providecommand{\URLprefix}{URL: }
\providecommand{\Pubmedprefix}{pmid:}
\providecommand{\doi}[1]{\href{http://dx.doi.org/#1}{\path{#1}}}
\providecommand{\Pubmed}[1]{\href{pmid:#1}{\path{#1}}}
\providecommand{\bibinfo}[2]{#2}
\ifx\xfnm\relax \def\xfnm[#1]{\unskip,\space#1}\fi
\bibitem[{Knapp and Carter(1976)}]{knapp}
\bibinfo{author}{C.~Knapp}, \bibinfo{author}{G.~Carter},
\newblock \bibinfo{title}{The generalized correlation method for estimation of
  time delay},
\newblock \bibinfo{journal}{IEEE Trans. Audio, Speech Lang. Process.}
  \bibinfo{volume}{24} (\bibinfo{year}{1976}) \bibinfo{pages}{320--327}.
\bibitem[{Carter(1987)}]{carter}
\bibinfo{author}{G.~C. Carter},
\newblock \bibinfo{title}{Coherence and time delay estimation},
\newblock \bibinfo{journal}{Proc. IEEE} \bibinfo{volume}{75}
  (\bibinfo{year}{1987}) \bibinfo{pages}{236--255}.
\bibitem[{Amoiridis et~al.(2022)Amoiridis, Zarri, Zamponi, Pasco, Yakhina,
  Christophe, Moreau, and Schram}]{AMOIRIDIS2022116534}
\bibinfo{author}{O.~Amoiridis}, \bibinfo{author}{A.~Zarri},
  \bibinfo{author}{R.~Zamponi}, \bibinfo{author}{Y.~Pasco},
  \bibinfo{author}{G.~Yakhina}, \bibinfo{author}{J.~Christophe},
  \bibinfo{author}{S.~Moreau}, \bibinfo{author}{C.~Schram},
\newblock \bibinfo{title}{Sound localization and quantification analysis of an
  automotive engine cooling module},
\newblock \bibinfo{journal}{J. Sound Vib.} \bibinfo{volume}{517}
  (\bibinfo{year}{2022}) \bibinfo{pages}{116534}.
\bibitem[{Martins et~al.(2023)Martins, Karimi, Maxit, and
  Kirby}]{MARTINS2023117891}
\bibinfo{author}{D.~Martins}, \bibinfo{author}{M.~Karimi},
  \bibinfo{author}{L.~Maxit}, \bibinfo{author}{R.~Kirby},
\newblock \bibinfo{title}{Non-negative intensity for a heavy fluid-loaded
  stiffened plate},
\newblock \bibinfo{journal}{J. Sound Vib.} \bibinfo{volume}{566}
  (\bibinfo{year}{2023}) \bibinfo{pages}{117891}.
\bibitem[{Grumiaux et~al.(2022)Grumiaux, Kitić, Girin, and
  Guérin}]{https://doi.org/10.48550/arxiv.2109.03465}
\bibinfo{author}{P.-A. Grumiaux}, \bibinfo{author}{S.~Kitić},
  \bibinfo{author}{L.~Girin}, \bibinfo{author}{A.~Guérin},
\newblock \bibinfo{title}{A survey of sound source localization with deep
  learning methods},
\newblock \bibinfo{journal}{J. Acoust. Soc. Am.} \bibinfo{volume}{152}
  (\bibinfo{year}{2022}) \bibinfo{pages}{107--151}.
\bibitem[{Gao et~al.(2023)Gao, Kuai, Huang, and Jiang}]{gao2023localization}
\bibinfo{author}{K.~Gao}, \bibinfo{author}{H.~Kuai},
  \bibinfo{author}{S.~Huang}, \bibinfo{author}{W.~Jiang},
\newblock \bibinfo{title}{Localization of broadband acoustical sources in the
  cylindrical duct via measurements outside the duct end},
\newblock \bibinfo{journal}{J. Sound Vib.}  (\bibinfo{year}{2023})
  \bibinfo{pages}{117749}.
\bibitem[{DiPassio et~al.(2023)DiPassio, Heilemann, and
  Bocko}]{DIPASSIO2023117671}
\bibinfo{author}{T.~DiPassio}, \bibinfo{author}{M.~C. Heilemann},
  \bibinfo{author}{M.~F. Bocko},
\newblock \bibinfo{title}{Direction of arrival estimation of an acoustic wave
  using a single structural vibration sensor},
\newblock \bibinfo{journal}{J. Sound Vib.} \bibinfo{volume}{553}
  (\bibinfo{year}{2023}) \bibinfo{pages}{117671}.
\bibitem[{Kita and Kajikawa(2021)}]{kita2021fundamental}
\bibinfo{author}{S.~Kita}, \bibinfo{author}{Y.~Kajikawa},
\newblock \bibinfo{title}{Fundamental study on sound source localization inside
  a structure using a deep neural network and computer-aided engineering},
\newblock \bibinfo{journal}{J. Sound Vib.} \bibinfo{volume}{513}
  (\bibinfo{year}{2021}) \bibinfo{pages}{116400}.
\bibitem[{Poschadel et~al.(2021)Poschadel, Hupke, Preihs, and
  Peissig}]{poschadel2021direction}
\bibinfo{author}{N.~Poschadel}, \bibinfo{author}{R.~Hupke},
  \bibinfo{author}{S.~Preihs}, \bibinfo{author}{J.~Peissig},
\newblock \bibinfo{title}{Direction of arrival estimation of noisy speech using
  convolutional recurrent neural networks with higher-order ambisonics
  signals},
\newblock in: \bibinfo{booktitle}{EUSIPCO}, \bibinfo{year}{2021}, pp.
  \bibinfo{pages}{211--215}.
\bibitem[{He et~al.(2019)He, Motlicek, and Odobez}]{he2019adaptation}
\bibinfo{author}{W.~He}, \bibinfo{author}{P.~Motlicek}, \bibinfo{author}{J.-M.
  Odobez},
\newblock \bibinfo{title}{Adaptation of multiple sound source localization
  neural networks with weak supervision and domain-adversarial training},
\newblock in: \bibinfo{booktitle}{Proc. IEEE Int. Conf. Acoust. Speech Signal
  Process.}, \bibinfo{year}{2019}, pp. \bibinfo{pages}{770--774}.
\bibitem[{Takeda and Komatani(2017)}]{takeda2017unsupervised}
\bibinfo{author}{R.~Takeda}, \bibinfo{author}{K.~Komatani},
\newblock \bibinfo{title}{Unsupervised adaptation of deep neural networks for
  sound source localization using entropy minimization},
\newblock in: \bibinfo{booktitle}{Proc. IEEE Int. Conf. Acoust. Speech Signal
  Process.}, \bibinfo{year}{2017}, pp. \bibinfo{pages}{2217--2221}.
\bibitem[{Takeda et~al.(2018)Takeda, Kudo, Takashima, Kitamura, and
  Komatani}]{takeda2018unsupervised}
\bibinfo{author}{R.~Takeda}, \bibinfo{author}{Y.~Kudo},
  \bibinfo{author}{K.~Takashima}, \bibinfo{author}{Y.~Kitamura},
  \bibinfo{author}{K.~Komatani},
\newblock \bibinfo{title}{Unsupervised adaptation of neural networks for
  discriminative sound source localization with eliminative constraint},
\newblock in: \bibinfo{booktitle}{Proc. IEEE Int. Conf. Acoust. Speech Signal
  Process.}, \bibinfo{year}{2018}, pp. \bibinfo{pages}{3514--3518}.
\bibitem[{He et~al.(2021)He, Motlicek, and Odobez}]{he2021neural}
\bibinfo{author}{W.~He}, \bibinfo{author}{P.~Motlicek}, \bibinfo{author}{J.-M.
  Odobez},
\newblock \bibinfo{title}{Neural network adaptation and data augmentation for
  multi-speaker direction-of-arrival estimation},
\newblock \bibinfo{journal}{IEEE Trans. Audio, Speech Lang. Process.}
  \bibinfo{volume}{29} (\bibinfo{year}{2021}) \bibinfo{pages}{1303--1317}.
\bibitem[{Shimodaira(2000)}]{shimodaira2000improving}
\bibinfo{author}{H.~Shimodaira},
\newblock \bibinfo{title}{Improving predictive inference under covariate shift
  by weighting the log-likelihood function},
\newblock \bibinfo{journal}{J. Stat. Plan.} \bibinfo{volume}{90}
  (\bibinfo{year}{2000}) \bibinfo{pages}{227--244}.
\bibitem[{Moreno-Torres et~al.(2012)Moreno-Torres, Raeder,
  Alaiz-Rodr{\'\i}guez, Chawla, and Herrera}]{moreno2012unifying}
\bibinfo{author}{J.~G. Moreno-Torres}, \bibinfo{author}{T.~Raeder},
  \bibinfo{author}{R.~Alaiz-Rodr{\'\i}guez}, \bibinfo{author}{N.~V. Chawla},
  \bibinfo{author}{F.~Herrera},
\newblock \bibinfo{title}{A unifying view on dataset shift in classification},
\newblock \bibinfo{journal}{Pattern Recognit.} \bibinfo{volume}{45}
  (\bibinfo{year}{2012}) \bibinfo{pages}{521--530}.
\bibitem[{Quiñonero-Candela et~al.(2008)Quiñonero-Candela, Sugiyama,
  Schwaighofer, and Lawrence}]{datashift}
\bibinfo{author}{J.~Quiñonero-Candela}, \bibinfo{author}{M.~Sugiyama},
  \bibinfo{author}{A.~Schwaighofer}, \bibinfo{author}{N.~D. Lawrence},
  \bibinfo{title}{Dataset Shift in Machine Learning}, \bibinfo{publisher}{The
  MIT Press}, \bibinfo{year}{2008}.
  \DOIprefix\doi{10.7551/mitpress/9780262170055.001.0001}.
\bibitem[{Tan et~al.(2018)Tan, Sun, Kong, Zhang, Yang, and Liu}]{tan2018survey}
\bibinfo{author}{C.~Tan}, \bibinfo{author}{F.~Sun}, \bibinfo{author}{T.~Kong},
  \bibinfo{author}{W.~Zhang}, \bibinfo{author}{C.~Yang},
  \bibinfo{author}{C.~Liu},
\newblock \bibinfo{title}{A survey on deep transfer learning},
\newblock in: \bibinfo{booktitle}{ICANN}, \bibinfo{year}{2018}, pp.
  \bibinfo{pages}{270--279}.
\bibitem[{Wang and Deng(2018)}]{wang2018deep}
\bibinfo{author}{M.~Wang}, \bibinfo{author}{W.~Deng},
\newblock \bibinfo{title}{Deep visual domain adaptation: A survey},
\newblock \bibinfo{journal}{Neurocomputing} \bibinfo{volume}{312}
  (\bibinfo{year}{2018}) \bibinfo{pages}{135--153}.
\bibitem[{Tzeng et~al.(2017)Tzeng, Hoffman, Saenko, and
  Darrell}]{tzeng2017adversarial}
\bibinfo{author}{E.~Tzeng}, \bibinfo{author}{J.~Hoffman},
  \bibinfo{author}{K.~Saenko}, \bibinfo{author}{T.~Darrell},
\newblock \bibinfo{title}{Adversarial discriminative domain adaptation},
\newblock in: \bibinfo{booktitle}{CVPR}, \bibinfo{year}{2017}, pp.
  \bibinfo{pages}{7167--7176}.
\bibitem[{Hoffman et~al.(2018)Hoffman, Tzeng, Park, Zhu, Isola, Saenko, Efros,
  and Darrell}]{hoffman2018cycada}
\bibinfo{author}{J.~Hoffman}, \bibinfo{author}{E.~Tzeng},
  \bibinfo{author}{T.~Park}, \bibinfo{author}{J.-Y. Zhu},
  \bibinfo{author}{P.~Isola}, \bibinfo{author}{K.~Saenko},
  \bibinfo{author}{A.~Efros}, \bibinfo{author}{T.~Darrell},
\newblock \bibinfo{title}{Cycada: Cycle-consistent adversarial domain
  adaptation},
\newblock in: \bibinfo{booktitle}{ICML}, \bibinfo{year}{2018}, pp.
  \bibinfo{pages}{1989--1998}.
\bibitem[{Saito et~al.(2018)Saito, Watanabe, Ushiku, and
  Harada}]{saito2018maximum}
\bibinfo{author}{K.~Saito}, \bibinfo{author}{K.~Watanabe},
  \bibinfo{author}{Y.~Ushiku}, \bibinfo{author}{T.~Harada},
\newblock \bibinfo{title}{Maximum classifier discrepancy for unsupervised
  domain adaptation},
\newblock in: \bibinfo{booktitle}{CVPR}, \bibinfo{year}{2018}, pp.
  \bibinfo{pages}{3723--3732}.
\bibitem[{Ajakan et~al.(2014)Ajakan, Germain, Larochelle, Laviolette, and
  Marchand}]{ajakan2014domain}
\bibinfo{author}{H.~Ajakan}, \bibinfo{author}{P.~Germain},
  \bibinfo{author}{H.~Larochelle}, \bibinfo{author}{F.~Laviolette},
  \bibinfo{author}{M.~Marchand},
\newblock \bibinfo{title}{Domain-adversarial neural networks},
\newblock \bibinfo{journal}{arXiv preprint arXiv:1412.4446}
  (\bibinfo{year}{2014}).
\bibitem[{Laradji and Babanezhad(2018)}]{laradji2020m}
\bibinfo{author}{I.~H. Laradji}, \bibinfo{author}{R.~Babanezhad},
\newblock \bibinfo{title}{M-adda: Unsupervised domain adaptation with deep
  metric learning},
\newblock \bibinfo{journal}{arXiv preprint arXiv:1807.02552}
  (\bibinfo{year}{2018}).
\bibitem[{Wilson and Cook(2020)}]{wilson2020survey}
\bibinfo{author}{G.~Wilson}, \bibinfo{author}{D.~J. Cook},
\newblock \bibinfo{title}{A survey of unsupervised deep domain adaptation},
\newblock \bibinfo{journal}{ACM Trans. Intell. Syst. Technol.}
  \bibinfo{volume}{11} (\bibinfo{year}{2020}) \bibinfo{pages}{1--46}.
\bibitem[{Kita and Kajikawa(2023)}]{kita_semi}
\bibinfo{author}{S.~Kita}, \bibinfo{author}{Y.~Kajikawa},
\newblock \bibinfo{title}{Study on sound source localization inside a structure
  using a domain transfer model for real-world adaption of a trained model},
\newblock in: \bibinfo{booktitle}{INTER-NOISE and NOISE-CON Congress and
  Conference Proceedings}, \bibinfo{number}{6},
  \bibinfo{organization}{Institute of Noise Control Engineering},
  \bibinfo{year}{2023}, pp. \bibinfo{pages}{1239--1248}.
\bibitem[{Zhu et~al.(2017)Zhu, Park, Isola, and Efros}]{zhu2017unpaired}
\bibinfo{author}{J.-Y. Zhu}, \bibinfo{author}{T.~Park},
  \bibinfo{author}{P.~Isola}, \bibinfo{author}{A.~A. Efros},
\newblock \bibinfo{title}{Unpaired image-to-image translation using
  cycle-consistent adversarial networks},
\newblock in: \bibinfo{booktitle}{ICCV}, \bibinfo{year}{2017}, pp.
  \bibinfo{pages}{2223--2232}.
\bibitem[{Isola et~al.(2017)Isola, Zhu, Zhou, and Efros}]{isola2017image}
\bibinfo{author}{P.~Isola}, \bibinfo{author}{J.-Y. Zhu},
  \bibinfo{author}{T.~Zhou}, \bibinfo{author}{A.~A. Efros},
\newblock \bibinfo{title}{Image-to-image translation with conditional
  adversarial networks},
\newblock in: \bibinfo{booktitle}{CVPR}, \bibinfo{year}{2017}, pp.
  \bibinfo{pages}{1125--1134}.
\bibitem[{Goodfellow et~al.(2014)Goodfellow, Pouget-Abadie, Mirza, Xu,
  Warde-Farley, Ozair, Courville, and Bengio}]{goodfellow2014generative}
\bibinfo{author}{I.~Goodfellow}, \bibinfo{author}{J.~Pouget-Abadie},
  \bibinfo{author}{M.~Mirza}, \bibinfo{author}{B.~Xu},
  \bibinfo{author}{D.~Warde-Farley}, \bibinfo{author}{S.~Ozair},
  \bibinfo{author}{A.~Courville}, \bibinfo{author}{Y.~Bengio},
\newblock \bibinfo{title}{Generative adversarial nets},
\newblock \bibinfo{journal}{NeurIPS} \bibinfo{volume}{27}
  (\bibinfo{year}{2014}).
\bibitem[{Mirza and Osindero(2014)}]{mirza2014conditional}
\bibinfo{author}{M.~Mirza}, \bibinfo{author}{S.~Osindero},
\newblock \bibinfo{title}{Conditional generative adversarial nets},
\newblock \bibinfo{journal}{arXiv preprint arXiv:1411.1784}
  (\bibinfo{year}{2014}).
\bibitem[{Yu et~al.(2019)Yu, Han, Shan, Dantcheva, and Chen}]{yu2019improving}
\bibinfo{author}{S.~Yu}, \bibinfo{author}{H.~Han}, \bibinfo{author}{S.~Shan},
  \bibinfo{author}{A.~Dantcheva}, \bibinfo{author}{X.~Chen},
\newblock \bibinfo{title}{Improving face sketch recognition via adversarial
  sketch-photo transformation},
\newblock in: \bibinfo{booktitle}{FG}, \bibinfo{year}{2019}, pp.
  \bibinfo{pages}{1--8}.
\bibitem[{Tang et~al.(2019)Tang, Wang, Wu, Chen, Xu, Sebe, and
  Yan}]{tang2019expression}
\bibinfo{author}{H.~Tang}, \bibinfo{author}{W.~Wang}, \bibinfo{author}{S.~Wu},
  \bibinfo{author}{X.~Chen}, \bibinfo{author}{D.~Xu},
  \bibinfo{author}{N.~Sebe}, \bibinfo{author}{Y.~Yan},
\newblock \bibinfo{title}{Expression conditional gan for facial
  expression-to-expression translation},
\newblock in: \bibinfo{booktitle}{ICIP}, \bibinfo{year}{2019}, pp.
  \bibinfo{pages}{4449--4453}.
\bibitem[{Yook et~al.(2018)Yook, Yoo, and Yoo}]{yook2018voice}
\bibinfo{author}{D.~Yook}, \bibinfo{author}{I.-C. Yoo},
  \bibinfo{author}{S.~Yoo},
\newblock \bibinfo{title}{Voice conversion using conditional cyclegan},
\newblock in: \bibinfo{booktitle}{CSCI}, \bibinfo{year}{2018}, pp.
  \bibinfo{pages}{1460--1461}.
\bibitem[{Lee et~al.(2020)Lee, Ko, Lee, Yoo, and Yook}]{lee2020many}
\bibinfo{author}{S.~Lee}, \bibinfo{author}{B.~Ko}, \bibinfo{author}{K.~Lee},
  \bibinfo{author}{I.-C. Yoo}, \bibinfo{author}{D.~Yook},
\newblock \bibinfo{title}{Many-to-many voice conversion using conditional
  cycle-consistent adversarial networks},
\newblock in: \bibinfo{booktitle}{Proc. IEEE Int. Conf. Acoust. Speech Signal
  Process.}, \bibinfo{year}{2020}, pp. \bibinfo{pages}{6279--6283}.
\bibitem[{Odena et~al.(2017)Odena, Olah, and Shlens}]{odena2017conditional}
\bibinfo{author}{A.~Odena}, \bibinfo{author}{C.~Olah},
  \bibinfo{author}{J.~Shlens},
\newblock \bibinfo{title}{Conditional image synthesis with auxiliary classifier
  gans},
\newblock in: \bibinfo{booktitle}{ICML}, \bibinfo{year}{2017}, pp.
  \bibinfo{pages}{2642--2651}.
\bibitem[{Naritomi et~al.(2018)Naritomi, Tanno, Ege, and
  Yanai}]{naritomi2018foodchangelens}
\bibinfo{author}{S.~Naritomi}, \bibinfo{author}{R.~Tanno},
  \bibinfo{author}{T.~Ege}, \bibinfo{author}{K.~Yanai},
\newblock \bibinfo{title}{Foodchangelens: Cnn-based food transformation on
  hololens},
\newblock in: \bibinfo{booktitle}{AIVR}, \bibinfo{year}{2018}, pp.
  \bibinfo{pages}{197--199}.
\bibitem[{Horita et~al.(2018)Horita, Tanno, Shimoda, and
  Yanai}]{horita2018food}
\bibinfo{author}{D.~Horita}, \bibinfo{author}{R.~Tanno},
  \bibinfo{author}{W.~Shimoda}, \bibinfo{author}{K.~Yanai},
\newblock \bibinfo{title}{Food category transfer with conditional cyclegan and
  a large-scale food image dataset},
\newblock in: \bibinfo{booktitle}{MADiMa}, \bibinfo{year}{2018}, pp.
  \bibinfo{pages}{67--70}.
\bibitem[{Bozorgtabar et~al.(2019)Bozorgtabar, Rad, Ekenel, and
  Thiran}]{bozorgtabar2019using}
\bibinfo{author}{B.~Bozorgtabar}, \bibinfo{author}{M.~S. Rad},
  \bibinfo{author}{H.~K. Ekenel}, \bibinfo{author}{J.-P. Thiran},
\newblock \bibinfo{title}{Using photorealistic face synthesis and domain
  adaptation to improve facial expression analysis},
\newblock in: \bibinfo{booktitle}{FG}, \bibinfo{year}{2019}, pp.
  \bibinfo{pages}{1--8}.
\bibitem[{Choi et~al.(2018)Choi, Choi, Kim, Ha, Kim, and
  Choo}]{choi2018stargan}
\bibinfo{author}{Y.~Choi}, \bibinfo{author}{M.~Choi}, \bibinfo{author}{M.~Kim},
  \bibinfo{author}{J.-W. Ha}, \bibinfo{author}{S.~Kim},
  \bibinfo{author}{J.~Choo},
\newblock \bibinfo{title}{Stargan: Unified generative adversarial networks for
  multi-domain image-to-image translation},
\newblock in: \bibinfo{booktitle}{CVPR}, \bibinfo{year}{2018}, pp.
  \bibinfo{pages}{8789--8797}.
\bibitem[{Mao et~al.(2017)Mao, Li, Xie, Lau, Wang, and
  Paul~Smolley}]{mao2017least}
\bibinfo{author}{X.~Mao}, \bibinfo{author}{Q.~Li}, \bibinfo{author}{H.~Xie},
  \bibinfo{author}{R.~Y. Lau}, \bibinfo{author}{Z.~Wang},
  \bibinfo{author}{S.~Paul~Smolley},
\newblock \bibinfo{title}{Least squares generative adversarial networks},
\newblock in: \bibinfo{booktitle}{ICCV}, \bibinfo{year}{2017}, pp.
  \bibinfo{pages}{2794--2802}.
\bibitem[{Inc.(2017)}]{Ansys}
\bibinfo{author}{A.~Inc.}, \bibinfo{title}{Mechanical apdl theory reference,
  release 18.2, 256--258.}, \bibinfo{year}{2017}. \URLprefix
  \url{https://www.mm.bme.hu/~gyebro/files/ans_help_v182/ans_thry/thy_acou2.html},
  \bibinfo{note}{(accessed 11 August 2021)}.
\bibitem[{Dare(2020)}]{70216d058af64a5f98c208ef90894204}
\bibinfo{author}{T.~Dare},
\newblock \bibinfo{title}{Experimental force reconstruction using a neural
  network and simulated training data},
\newblock in: \bibinfo{booktitle}{INTER-NOISE}, \bibinfo{year}{2020}, pp.
  \bibinfo{pages}{4995--5868}.
\bibitem[{Dare(2021)}]{6d68305e44514951b940a0223d30a0af}
\bibinfo{author}{T.~Dare},
\newblock \bibinfo{title}{Experimental force reconstruction on plates of
  arbitrary shape using neural networks},
\newblock in: \bibinfo{booktitle}{INTER-NOISE}, \bibinfo{year}{2021}, pp.
  \bibinfo{pages}{2949--3943}. \DOIprefix\doi{10.3397/IN2021-2397}.
\bibitem[{He et~al.(2016)He, Zhang, Ren, and Sun}]{he2016deep}
\bibinfo{author}{K.~He}, \bibinfo{author}{X.~Zhang}, \bibinfo{author}{S.~Ren},
  \bibinfo{author}{J.~Sun},
\newblock \bibinfo{title}{Deep residual learning for image recognition},
\newblock in: \bibinfo{booktitle}{CVPR}, \bibinfo{year}{2016}, pp.
  \bibinfo{pages}{770--778}.
\bibitem[{Zeiler et~al.(2010)Zeiler, Krishnan, Taylor, and
  Fergus}]{zeiler2010deconvolutional}
\bibinfo{author}{M.~D. Zeiler}, \bibinfo{author}{D.~Krishnan},
  \bibinfo{author}{G.~W. Taylor}, \bibinfo{author}{R.~Fergus},
\newblock \bibinfo{title}{Deconvolutional networks},
\newblock in: \bibinfo{booktitle}{CVPR}, \bibinfo{year}{2010}, pp.
  \bibinfo{pages}{2528--2535}.
\bibitem[{Ulyanov et~al.(2016)Ulyanov, Vedaldi, and
  Lempitsky}]{ulyanov2016instance}
\bibinfo{author}{D.~Ulyanov}, \bibinfo{author}{A.~Vedaldi},
  \bibinfo{author}{V.~Lempitsky},
\newblock \bibinfo{title}{Instance normalization: The missing ingredient for
  fast stylization},
\newblock \bibinfo{journal}{arXiv preprint arXiv:1607.08022}
  (\bibinfo{year}{2016}).
\bibitem[{Maas et~al.(2013)Maas, Hannun, Ng et~al.}]{maas2013rectifier}
\bibinfo{author}{A.~L. Maas}, \bibinfo{author}{A.~Y. Hannun},
  \bibinfo{author}{A.~Y. Ng}, et~al.,
\newblock \bibinfo{title}{Rectifier nonlinearities improve neural network
  acoustic models},
\newblock in: \bibinfo{booktitle}{ICML}, volume~\bibinfo{volume}{30},
  \bibinfo{year}{2013}, p.~\bibinfo{pages}{3}.
\bibitem[{Zhong et~al.(2020)Zhong, Zheng, Kang, Li, and Yang}]{zhong2020random}
\bibinfo{author}{Z.~Zhong}, \bibinfo{author}{L.~Zheng},
  \bibinfo{author}{G.~Kang}, \bibinfo{author}{S.~Li},
  \bibinfo{author}{Y.~Yang},
\newblock \bibinfo{title}{Random erasing data augmentation},
\newblock in: \bibinfo{booktitle}{AAAI}, volume~\bibinfo{volume}{34},
  \bibinfo{year}{2020}, pp. \bibinfo{pages}{13001--13008}.
\bibitem[{DeVries and Taylor(2017)}]{devries2017improved}
\bibinfo{author}{T.~DeVries}, \bibinfo{author}{G.~W. Taylor},
\newblock \bibinfo{title}{Improved regularization of convolutional neural
  networks with cutout},
\newblock \bibinfo{journal}{arXiv preprint arXiv:1708.04552}
  (\bibinfo{year}{2017}).
\bibitem[{Van~der Maaten and Hinton(2008)}]{van2008visualizing}
\bibinfo{author}{L.~Van~der Maaten}, \bibinfo{author}{G.~Hinton},
\newblock \bibinfo{title}{Visualizing data using t-sne.},
\newblock \bibinfo{journal}{JMLR} \bibinfo{volume}{9} (\bibinfo{year}{2008}).
\bibitem[{Selvaraju et~al.(2017)Selvaraju, Cogswell, Das, Vedantam, Parikh, and
  Batra}]{selvaraju2017grad}
\bibinfo{author}{R.~R. Selvaraju}, \bibinfo{author}{M.~Cogswell},
  \bibinfo{author}{A.~Das}, \bibinfo{author}{R.~Vedantam},
  \bibinfo{author}{D.~Parikh}, \bibinfo{author}{D.~Batra},
\newblock \bibinfo{title}{Grad-cam: Visual explanations from deep networks via
  gradient-based localization},
\newblock in: \bibinfo{booktitle}{ICCV}, \bibinfo{year}{2017}, pp.
  \bibinfo{pages}{618--626}.
\bibitem[{Kita and Kajikawa(2023)}]{kita2023}
\bibinfo{author}{S.~Kita}, \bibinfo{author}{Y.~Kajikawa},
\newblock \bibinfo{title}{Sound source localization inside a structure under
  semi-supervised conditions},
\newblock \bibinfo{journal}{IEEE/ACM Transactions on Audio, Speech, and
  Language Processing} \bibinfo{volume}{31} (\bibinfo{year}{2023})
  \bibinfo{pages}{1397--1408}.
\bibitem[{Parzen(1962)}]{parzen1962estimation}
\bibinfo{author}{E.~Parzen},
\newblock \bibinfo{title}{On estimation of a probability density function and
  mode},
\newblock \bibinfo{journal}{Ann. Math. Stat.} \bibinfo{volume}{33}
  (\bibinfo{year}{1962}) \bibinfo{pages}{1065--1076}.

\end{thebibliography}

\bibliographystyle{model1-num-names}







\end{document}